\documentclass[preprint]{elsarticle}
\pdfoutput=1

\usepackage{natbib}
\usepackage{lineno}

\title{Cosmic Ray Muon Flux at the Sanford Underground Laboratory at Homestake}

\author[regis,mines]{F.E.~Gray\corref{cor1}}
\author[regis]{C.~Ruybal\fnref{ruybal}}
\fntext[ruybal]{Present address: Colorado School of Mines, Environmental Science and Engineering Division, 1500 Illinois St., Golden, CO 80401, U.S.A.}
\author[regis]{J.~Totushek\fnref{totushek}}
\fntext[totushek]{Present address: North Dakota State University, Department of Mathematics, 1210 Albrecht Boulevard, Fargo, ND  58102, U.S.A.}
\author[usd]{D.-M.~Mei}
\author[usd]{K.~Thomas}
\author[usd,china]{C.~Zhang}

\address[regis]{Regis University, Department of Physics and Computational Science,
  3333 Regis Blvd., Denver, CO  80221, U.S.A.}
\address[mines]{Colorado School of Mines, Department of Physics, 1523 Illinois St., Golden, CO  80401, U.S.A.}
\address[usd]{University of South Dakota, Department of Physics,
  414 E. Clark St.,  Vermillion, SD  57069, U.S.A.}
\address[china]{China Three Gorges University, College of Science, Yichang 443002, China}

\cortext[cor1]{Corresponding author}

\begin{document}


\begin{abstract}
\noindent 

Measuring the muon flux is important to the Sanford Underground Laboratory at Homestake, for which several low background
experiments are being planned.  The nearly-vertical cosmic ray muon flux was measured
in three locations at this laboratory: on the surface ($1.149\pm0.017 \times 10^{-2}~\rm{cm}^{-2}~\rm{s}^{-1}~\rm{sr}^{-1}$), at the 800-ft (0.712 km w.e.) level ($2.67\pm0.06 \times 10^{-6}~\rm{cm}^{-2}~\rm{s}^{-1}~\rm{sr}^{-1}$),
and at the 2000-ft (1.78 km w.e.) level ($2.56\pm0.25 \times 10^{-7}~\rm{cm}^{-2}~\rm{s}^{-1}~\rm{sr}^{-1}$).  These fluxes agree well with model predictions. 

\noindent
\end{abstract}

\begin{keyword}
muon flux, underground laboratory

\PACS 29.90.+r, 95.45.+i, 95.55.Vj
\end{keyword}

\maketitle


The Homestake Mine in Lead, South Dakota, USA was identified in 2007 as the 
final candidate site for the Deep Underground Science and Engineering 
Laboratory (DUSEL).  In advance of the federal funds to further develop the 
mine, the South Dakota Science and Technology Authority (SDSTA), which operates
the Sanford Underground Laboratory, is offering an early science program mainly to characterize aspects of the site environment.
It is located at 44.35$^\circ$~N, 103.77$^\circ$~W, with a surface elevation 
of 1620~m above sea level.  
Initially, the LUX (Large Underground Xenon) dark matter 
search~\cite{Fiorucci:2009ak} and the {\sc{Majorana}} neutrinoless double beta 
decay experiment~\cite{Henning:2009tt} will be located there. 
Measurements of external backgrounds, including cosmic ray muons as well as 
gammas and neutrons, will be of paramount importance to the design of these 
sensitive rare-event searches.


The differential muon flux at the 4850 ft. (4.40~km water equivalent) level 
at Homestake was measured by Cherry $et$ $al$~\cite{mlc}.  However, more 
measurements are needed to characterize the muon flux as a function of depth.
This paper describes new  measurements of the cosmic ray muon flux at three 
locations: the surface (in a building that provided $\sim$1 m w.e. of 
shielding), the 800~ft.~(0.712 km w.e.) level, and the 
2000~ft.~(1.78 km w.e.) level.  We present a comparison between our 
reported flux and the model of~\cite{meihime}, and we find agreement between 
our experimental results and that model.

Cosmic-ray muon flux as a function of depth has been studied by many 
experiments at underground facilities around the world.  Measurements of the 
muon flux per unit solid angle as a function of slant depth from 
Castagnoli~\cite{cca}, Barrett~\cite{phb}, Miyake~\cite{smi}, 
WIPP~\cite{eie}, Soudan~\cite{kasa}, Kamioka~\cite{kamland}, 
Boulby~\cite{mro}, Gran Sasso~\cite{lvdd,macroo}, Fr\'{e}jus~\cite{chb} 
and Sudbury~\cite{PhysRevD.80.012001} have been used to develop a 
model, which can be used to predict the muon flux per unit solid angle 
as a function of depth~\cite{meihime}. 

\section{Methods}

\begin{figure}
\begin{center}
\includegraphics[height=6cm]{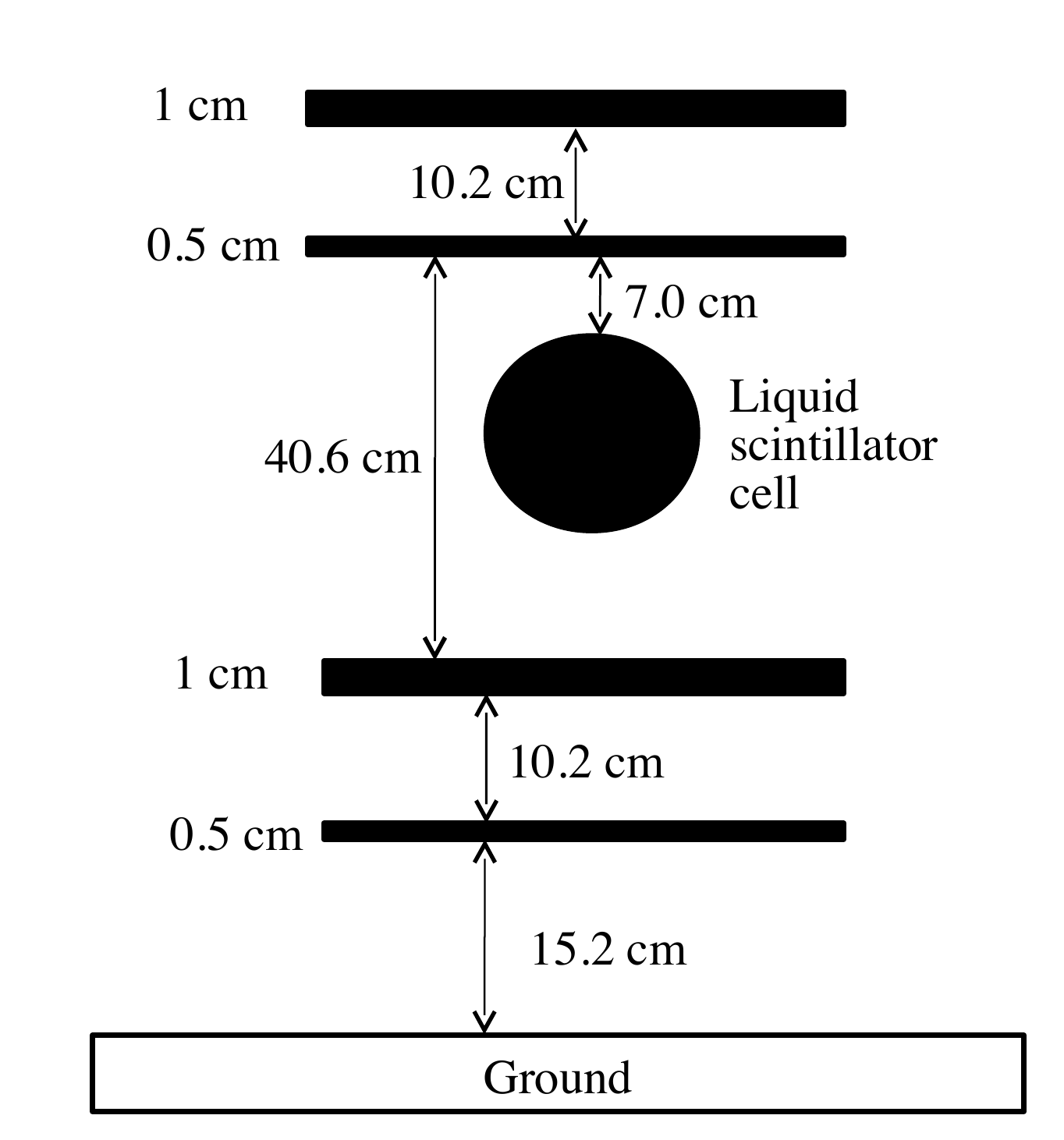}
\end{center}
\caption{Vertical positioning of elements of the muon detector.}
\label{fig:geometry}
\end{figure}

The muon detector consists of four fast plastic scintillation counters (Saint-Gobain BC-408), each a square with a side length of 30.5$\pm$0.1~cm.  As shown in 
Figure~\ref{fig:geometry}, the distance from the top to the bottom counter 
is 64.0~cm.  Two of these counters are 0.5~cm thick, and the others 
are 1.0~cm thick.  
Each is coupled by a trapezoidal acrylic lightguide to a Photonis~XP2020 
photomultiplier with Photonis VD124K base.   Waveforms from each detector 
are recorded by a 12-bit flash analog-to-digital conversion module with a 
sampling frequency of 170 MHz; it filters the data onboard with field 
programmable gate arrays and transmits digitized pulses through an Ethernet 
interface to a standard personal computer.

The detector station also includes a 1.2~L liquid scintillation counter 
filled with Eljen Technologies EJ-301 or EJ-309 material.  
This counter has been used to study techniques for neutron counting in 
the underground environment; a future publication will quantify neutron 
backgrounds in the mine.

The gamma ray flux at each of the sites is at the level of 1~cm$^{-2}$~s$^{-1}$; 
the counting rate for background gamma events in each of the plastic 
scintillator detectors is therefore $\sim$1~kHz.  Nearly all of the 
gamma flux is at energies of 2.5~MeV or less, as was reported 
in~\cite{Mei:2009py}.  A substantial coincidence requirement is 
needed to distinguish cosmic ray muon events from the gamma background.  At the 
2000~ft. depth, a two-fold coincidence analysis was found to be dominated
by the gamma background, and at shallower depths they still represent a 
substantial correction.  Consequently, all of our results are based on 
studies where we require
that at least three of the four detectors record a pulse within a $\sim$70~ns time interval.
Because of the higher counting rate and reduced sensitivity to several
systematic uncertainties, we present the three-fold coincidence analysis as our primary 
result, and a four-fold coincidence analysis (where all four detectors are 
required to fire simultaneously) as a partially independent check.
In particular, the agreement between the three-fold and four-fold coincidence 
results demonstrates that secondary particles such as high-energy electrons 
produced by muon interactions do not affect the counting rate substantially
relative to the precision of this measurement.

\begin{figure}
\begin{center}
\includegraphics[height=6cm]{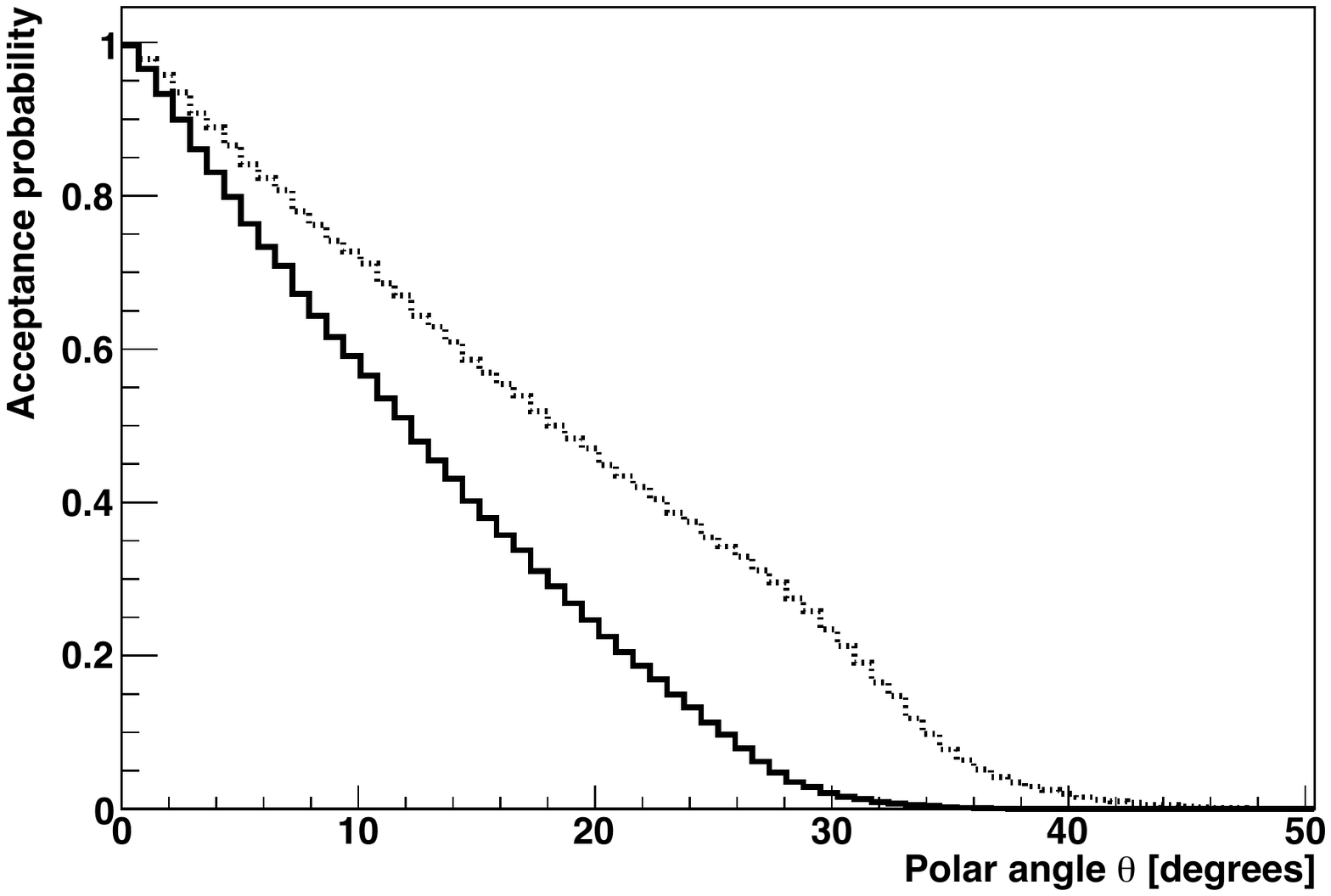}
\end{center}
\caption{Acceptance probability as a function of polar angle $\theta$ 
for the four-fold coincidence analysis (solid line) or the three-fold
coincidence analysis (dotted line).}
\label{fig:solidAngle}
\end{figure}

A geometric Monte Carlo calculation was used to determine the solid angle 
accepted by each of these analysis methods.  It took into account only the
size and position of each of the detector elements, assuming straight muon
tracks.   As shown in Figure~\ref{fig:solidAngle}, it was used to determine
the acceptance probability as a function of polar angle $P(\theta)$, which 
was then integrated to determine the accepted solid angle:
\begin{displaymath}
\Omega = 2\pi \int\limits_{0}^{\pi/2} P(\theta) \sin\theta \, d\theta  ~.
\end{displaymath}
When a four-fold coincidence is required, the accepted solid angle is
0.226~sr; it is 0.457~sr when only a three-fold coincidence is required. 
If we assume an incident muon distribution proportional to $\cos^2 \theta$, 
90\% of the flux in the three-fold coincidence analysis would be within
25$^\circ$ of vertical, while 90\% of the flux in the four-fold coincidence
analysis would be within 19$^\circ$ of vertical.

This Monte Carlo program was also used to study the systematic uncertainty 
arising from misalignment of the detector elements.  A horizontal displacement
of one detector element in this program by 2~cm, which is believed to
represent the worst realistic possibility for the data collected on the 800 ft. level, 
changed the calculated solid angle by a maximum of 0.4\% for the three-fold 
coincidence analysis and 1.1\% for the four-fold coincidence analysis.  
The alignment on other levels is believed to have been substantially better, 
with a maximum possible displacement of 1~cm.

The energy scales of the detectors were calibrated based on the observed
pulse amplitude spectra for four-fold coincident events.  All such events 
on the surface and at the 800 ft. level were presumed to be minimum-ionizing 
particles that would give a most probable energy deposition of 1.0~MeV in 
the thin detectors and 2.0~MeV in the thick detectors.  These energies were
computed from the scintillator density of 1.032 g/cm$^3$, assuming the 
minimum-ionizing $dE/dx$ given by the Bethe-Bloch equation~\cite{knoll}.
At the surface and 
the 800~ft. depth, there was sufficient statistical power to allow the 
calibration to be determined {\em in situ}.  
At the 2000~ft. depth, a 
calibration from the surface was applied; this method would have been 
preferred at the 800~ft. depth as well, but detector repairs required
recalibration to be completed underground.  

The digitization hardware thresholds 
were set as low as practical, corresponding to approximately 0.4~MeV for the 
thin detectors and 0.9~MeV for the thick detectors.  Analysis thresholds 
were then established at 0.75~MeV in the thin counters and 1.5~MeV in the thick counters.
These thresholds require a total energy deposition of at least 3~MeV for a
three-fold coincidence, which is beyond the endpoint of the gamma spectrum,
but still maintains an efficiency for muons that can be determined
effectively.

The efficiency corrections associated with these energy cuts were 
determined from the data.  At the surface, we assumed that all four-fold
coincident events were caused by minimum-ionizing particles; other particles 
from atmospheric showers would have been shielded effectively by 
the $\sim$1 m w.e.  provided by the building above the detector.  
Therefore, we computed the ratio of the number of events that passed the 
analysis cuts to the the total number of recorded four-fold coincidences 
where only the hardware thresholds were used.  

Having determined this efficiency on the surface, we then applied it to 
data collected at the 2000~ft. level.  The uncertainty associated with
this procedure includes a statistical component from the finite number of 
coincidences at the surface.  However, another part of the uncertainty is 
related to the stability of the detector gain, and therefore the energy scale, 
especially as the detector was being moved 
underground.  We checked for these gain shifts using three standard beta/gamma 
calibration sources, $^{22}$Na, $^{60}$Co, and $^{90}$Sr, collecting pulse
height spectra before and after relocating the detector.  Averaging the results
from these sources, we found gain reductions of 0.8\%, 2.0\%, 4.7\%, and 3.1\% 
for the four detector elements.  We corrected the energy scale in the analysis 
according to these results, and we treated the 1.0\% uncertainty in each 
detector calibration as the systematic error for the efficiency in the 
result from the 2000~ft. level.  Examination of the energy distribution 
in each detector showed that the single-detector rate would change by 0.4\% for
each percent change in the assumed energy scale, and a Monte Carlo simulation
verified the naive predictions that such a change would give a modification of
0.4\% to the three-fold coincidence analysis and 1.6\% to the four-fold coincidence
analysis.

Because it was necessary to recalibrate
the detector while it was at the 800~ft. level, a different procedure was 
applied there.
We determined separately for each scintillator the fraction 
of four-fold coincidence events where only the hardware threshold was 
imposed that would be cut by applying the energy threshold in that 
scintillator.  We incorporated these efficiencies into the geometric
Monte Carlo simulation program and observed the change in the number of 
accepted three-fold and four-fold coincidence events in order to determine the 
overall efficiency.  The uncertainty associated with this procedure is taken 
to be purely statistical; any systematic drift in the energy scale should 
present a negligible effect, as the detector remained untouched underground at 
nearly constant temperature.  The computed efficiency was substantially higher than before,
because an upgraded flash ADC electronics module was installed
when the detector was relocated to the 800~ft. level; the new module 
included an integrating filter that improved the effective energy resolution 
of the system.

The start and stop times of each data file run were determined from the 
acquisition computer's internal clock, which is synchronized using the 
Network Time Protocol with a stratum 2 server.  
These times were used for the primary determination of the experiment's 
live time.  In addition, the flash ADC module was set to generate periodic 
sampling triggers every 0.8~s, and these triggers were counted for a 
subset of the data as a 
cross-check and found to agree within 0.1\%.  The time base on the 
flash ADC board is derived from a crystal oscillator that is 
specified for accuracy and stability to $\pm$50~ppm.  

\section{Results}

We have determined that the rate of throughgoing muons on the surface,
on the lower floor of the SDSTA administration building ($\sim$1 m w.e.), is
\begin{displaymath}
(1.149\pm0.017) \times 10^{-2}~\rm{cm}^{-2}~\rm{s}^{-1}~\rm{sr}^{-1} ~.
\end{displaymath}
On the 800~ft. (0.712 km w.e.) level, in the former blasting cap storage area near the Ross 
shaft, the flux is
\begin{displaymath}
(2.67 \pm 0.06) \times 10^{-6}~\rm{cm}^{-2}~\rm{s}^{-1}~\rm{sr}^{-1} ~.
\end{displaymath}
Finally, on the 2000~ft. (1.78 km w.e.) level, at the first transformer pad site in 
the drift between the Ellison and Yates shafts, the flux is
\begin{displaymath}
(2.56\pm0.25) \times 10^{-7}~\rm{cm}^{-2}~\rm{s}^{-1}~\rm{sr}^{-1} ~.
\end{displaymath}
Table~\ref{tab:results} provides more details on the measurements that
led to these flux values, and it compares them to the corresponding 
results from the four-fold coincidence analysis method.
\begin{table}
\begin{small}
\begin{tabular}{llll}
\hline
{\bf Site/Level}  & {\bf Surface} & {\bf 800 ft.} & {\bf 2000 ft. } \\
\hline
Equivalent depth (km w.e.)  &  0.001    & 0.712 & 1.78 \\
\hline
Live time (s; $\pm$ 0.1\%) & 3794 & 6716383 & 1205647 \\
\hline
{\em Three-fold coincidence analysis:} & & \\
Number of events ($\pm \sqrt{N}$) &  15256 & 7138 & 108 \\
Efficiency of energy cut & 0.823 $\pm$ 0.009 & 0.938 $\pm$ 0.017 & 0.823 $\pm$ 0.010 \\
Solid angle (sr) & 0.457 $\pm$ 0.001 & 0.457 $\pm$ 0.002 & 0.457 $\pm$ 0.001 \\
\hline
Flux (cm$^{-2}$ s$^{-1}$ sr$^{-1}$) & (1.149 $\pm$ 0.017) $\times 10^{-2}$ & (2.67 $\pm$ 0.06) $\times 10^{-6}$ &  (2.56 $\pm$ 0.25) $\times 10^{-7}$ \\
\hline
\hline
{\em Four-fold coincidence analysis:} & & \\
Number of events ($\pm \sqrt{N}$) & 6930 & 3261 & 50 \\
Efficiency of energy cut & 0.734 $\pm$ 0.012  & 0.837 $\pm$ 0.021 & 0.734 $\pm$ 0.017 \\
Solid angle (sr) & 0.226 $\pm$ 0.001  & 0.226 $\pm$ 0.003 & 0.226 $\pm$ 0.001 \\
\hline
Flux (cm$^{-2}$ s$^{-1}$ sr$^{-1}$) & (1.184 $\pm$ 0.025) $\times 10^{-2}$ & (2.76 $\pm$ 0.09) $\times 10^{-6}$ & (2.69 $\pm$ 0.39) $\times 10^{-7}$ \\
\hline
\hline
\end{tabular}
\end{small}
\caption{Summary of results from different sites in the mine, showing the consistency of the three-fold and four-fold coincidence analysis methods.}
\label{tab:results}
\end{table}

\section{Conclusion}

\begin{figure}
\begin{center}
\includegraphics[height=8cm]{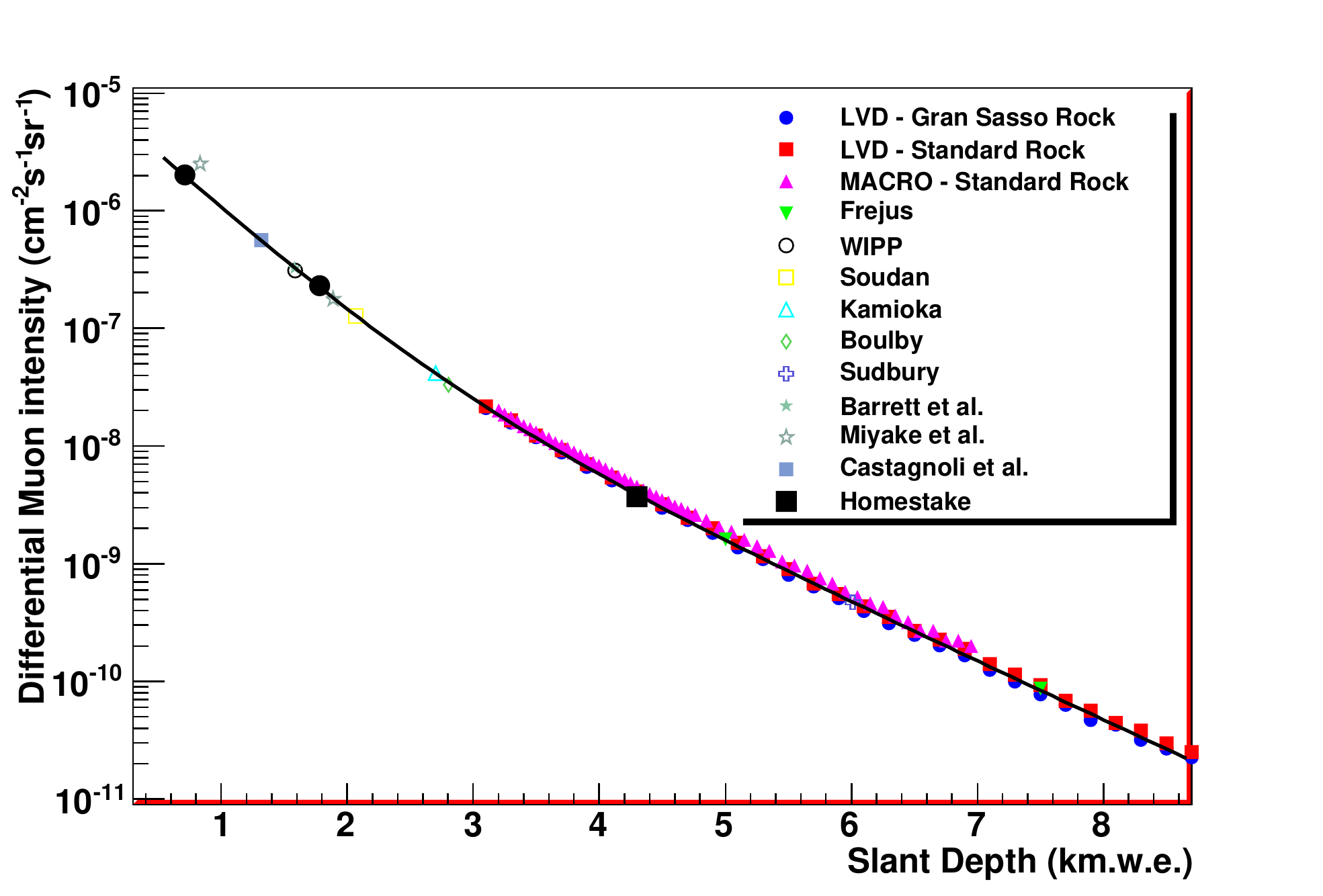}
\end{center}
\caption{Muon flux as a function of depth, placing these measurements in the context of previous results from Castagnoli~\cite{cca}, 
Barrett~\cite{phb},
Miyake~\cite{smi}, WIPP~\cite{eie}, Soudan~\cite{kasa}, Kamioka~\cite{kamland}, 
Boulby~\cite{mro}, Gran Sasso~\cite{lvdd}, Fr\'{e}jus~\cite{chb} and Sudbury~\cite{PhysRevD.80.012001}, and comparing them with a parameterization model~\cite{meihime}.  The three data points for Homestake are from this work (filled black circles) and Cherry~et~al.~\cite{mlc} (filled black square).}
\label{fig:com1}
\end{figure}

\begin{figure}
\begin{center}
\includegraphics[height=8cm]{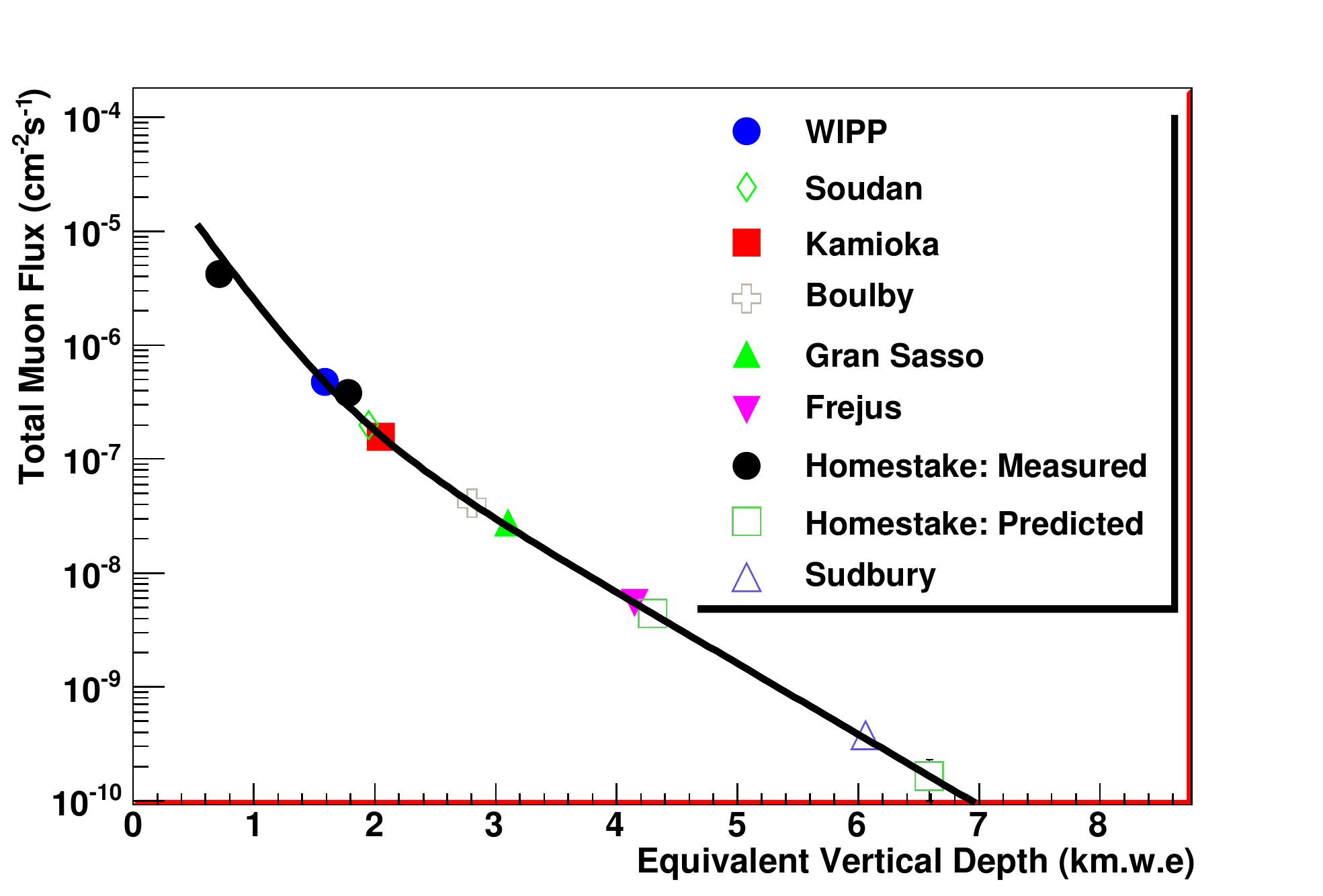}
\end{center}
\caption{Integrated muon flux as a function of depth, compared against a flat-earth model.  This model is 
used to extrapolate to the 4850 ft. and 8000 ft. levels at Homestake.}
\label{fig:com2}
\end{figure}

Figure~\ref{fig:com1} shows the muon flux as a function of the depth, 
comparing the measured fluxes to the prediction of a parameterization 
model~\cite{meihime}.  The agreement between the measurements and predictions 
for different levels is within 20\% in all cases.
The integrated muon fluxes at the different levels are also compared to the prediction of a flat-earth model~\cite{meihime}.
Figure~\ref{fig:com2} illustrates the agreement, which is clearly excellent over five orders of magnitude in flux.  
We have considered the effect of the large open cut in the mine; while it 
is quite deep, the slant depth to our experimental sites is large enough that 
it has a negligible effect on the flux.  Consequently, 
we have used these models to predict the differential and integral
fluxes at levels of scientific interest, 
as summarized in Table~\ref{tab:calc_fluxes}.
\begin{table}
\begin{tabular}{lll}
\hline
{\bf Site}  & {\bf Flux per unit } & {\bf Integral flux} \\
  & {\bf solid angle} & \\
  & ($\rm{cm}^{-2}~\rm{s}^{-1}~\rm{sr}^{-1}$) & ($\rm{cm}^{-2}~\rm{s}^{-1}$) \\
\hline
4850 ft. (1478~m) & $3.85 \times 10^{-9}$ & $4.40 \times 10^{-9}$ \\
7400 ft. (2255~m) & $2.21 \times 10^{-10}$ & $1.65 \times 10^{-10}$ \\
8000 ft. (2438~m) & $1.19 \times 10^{-10}$ & $6.86 \times 10^{-11}$ \\
\hline
\end{tabular}
\caption{Differential (per unit solid angle) and integral fluxes calculated 
at several levels in the Homestake Mine using modeling techniques described 
in~\cite{meihime}
that have been calibrated by the results described in this paper. 
Major experimental campuses are proposed for the 4850 ft. and 7400 ft. levels,
while 8000 ft. is the deepest level.}
\label{tab:calc_fluxes}
\end{table}

\section{Acknowledgments}

We thank Jaret Heise, Tom Trancynger, and all of the staff at the 
Sanford Laboratory for their valuable support.  We also thank Uwe Greife of
the Colorado School of Mines for allowing us to use his laboratory resources
for this project.  This work was supported by the National Science 
Foundation under grant NSF-PHY-0758120. 


\bibliographystyle{h-elsevier}

\bibliography{muon_flux}

\end{document}